\documentclass[prl,twocolumn,preprintnumbers]{revtex4-1} 

\usepackage{graphicx}
\usepackage{bm}
\usepackage{amsmath}
\usepackage{amssymb}
\usepackage{amsfonts}
\usepackage{fancyhdr}
\usepackage{slashed}
\usepackage{latexsym,epsfig,bbm}

\newcommand{\bra}[1]{\langle{#1}|}
\newcommand{\ket}[1]{|{#1}\rangle}

\newcommand{\realsum}{\displaystyle\sum}

\newcommand{\figref}[1]{Fig.\ \ref{#1}}
\newcommand{\figsref}[1]{Figs.\ \ref{#1}}
\def \d {\mathrm{d}}

\begin{document}

\title{Multipartite Entangled Spatial Modes of Ultracold Atoms Generated and Controlled by Quantum Measurement}

\author{T. J. Elliott}
\email{thomas.elliott@physics.ox.ac.uk}
\affiliation{Department of Physics, Clarendon Laboratory, University of Oxford, Parks Road, Oxford OX1 3PU, United Kingdom}
\author{W. Kozlowski}
\affiliation{Department of Physics, Clarendon Laboratory, University of Oxford, Parks Road, Oxford OX1 3PU, United Kingdom}
\author{S. F. Caballero-Benitez}
\affiliation{Department of Physics, Clarendon Laboratory, University of Oxford, Parks Road, Oxford OX1 3PU, United Kingdom}
\author{I. B. Mekhov}
\affiliation{Department of Physics, Clarendon Laboratory, University of Oxford, Parks Road, Oxford OX1 3PU, United Kingdom}

\date{\today}

\begin{abstract}
We show that the effect of measurement back-action results in the generation of multiple many-body spatial modes of ultracold atoms trapped in an optical lattice, when scattered light is detected. The multipartite mode entanglement properties and their nontrivial spatial overlap can be varied by tuning the optical geometry in a single setup. This can be used to engineer quantum states and dynamics of matter fields. We provide examples of multimode generalizations of parametric down-conversion, Dicke, and other states, investigate the entanglement properties of such states, and show how they can be transformed into a class of generalized squeezed states. Further, we propose how these modes can be used to detect and measure entanglement in quantum gases.
\end{abstract}
\maketitle

Recently, the field of quantum gases \cite{bloch2008,lewenstein2012} has grown considerably, due to the suitability of atomic systems for quantum simulation of a wide array of systems with origins in other fields, such as condensed matter and particle physics. Such systems also have use for entanglement and quantum information processing (QIP) \cite{monroe2002}. Control is achieved by light fields, and there has been recent interest in the regime when the light exhibits decidedly quantum properties, thus uniting quantum optics with many-body physics (see Refs.\ \cite{mekhov2012, ritsch2013} for reviews). This fully quantum regime enables one to go beyond standard questions of ultracold gases trapped in fixed classical potentials, thus broadening the field even further. 

Measurement back-action, the evolution of a state due to observation, is one of the primary manifestations of quantum mechanics. It was exploited in the breakthrough cavity QED experiments \cite{HarocheBook}, where atoms were used as probes of quantum states of light. Intriguing Fock and Schr{\"o}dinger cat states were prepared in a single cavity using quantum non-demolition (QND) methods. However, scaling to a large number of cavities provides an extreme challenge. 

In contrast, we consider a case where the roles of light and matter are reversed: ultracold atoms are trapped in an optical lattice, and light is used as a global QND probe. Thus, the lattice sites represent the storage of multiple quantum states of matter fields, and the number of illuminated sites can be tuned from few to thousands, enabling scaling. We show how the quantum nature of light manifest in the measurement back-action can be used to establish a rich mode structure of the matter fields, with nontrivial delocalization over many sites and entanglement properties. These modes can be used for quantum state engineering, including multimode generalizations of parametric down-conversion (PDC) and Dicke states. We focus on the mode entanglement properties of these states, which exhibit genuine multipartite mode entanglement \cite{ghirardi2004,amico2008}, and contrary to the entanglement inherent to the symmetrization of indistinguishable particles, may be extracted for use in QIP. In contrast to setups with atomic ensembles, we consider optical lattices, which enables the modes to have a significant amount of spatial overlap, and the light allows us to introduce effective long-range interactions, allowing for new schemes to be realized. Further, using their nontrivial spatial overlap, we suggest how they can be used in the measurement of entanglement. Being based on off-resonant scattering, these ideas can be exported to other systems, such as molecules \cite{mekhovLP2013}, fermions \cite{RuostekoskiFoot2009} and spins \cite{RuostekoskiAF2014}.

\begin{figure}
\centering
\includegraphics[width=0.9\linewidth]{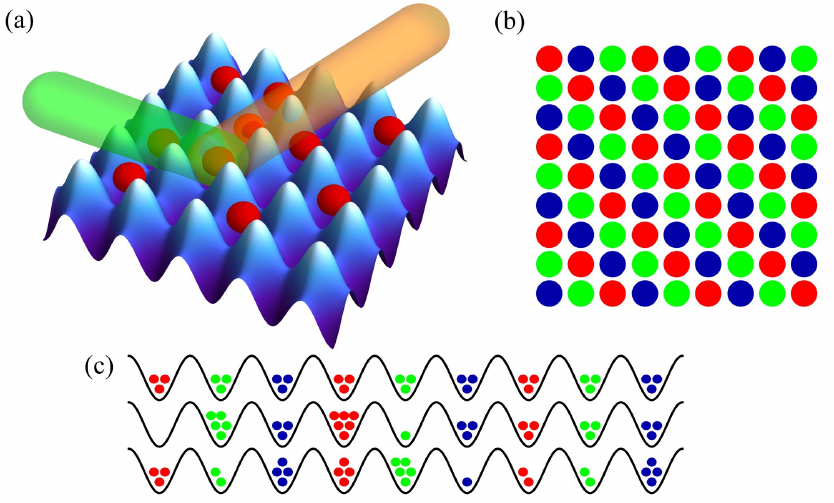}
\caption{(a) Setup: light is scattered from atoms in an optical lattice. Generated spatial structure of 3 matter-field modes in 2D (b) and 1D (c) lattice. Sites of the same colour are indistinguishable to light scattering and thus belong to the same mode. The superposition of three indistinguishable atom distributions in (c) is protected by light scattering.}
\label{figmodel}
\end{figure}

We study a generalized Bose-Hubbard model \cite{mekhov2012}, where atoms in an optical lattice scatter light [\figref{figmodel}(a)], and crucially, light is elevated to a dynamical variable (see Supplemental Material for details). As ultracold particles are delocalized, the atomic state is a superposition of Fock state configurations $\ket{\bm{n}}=\ket{n_1,..,n_M}$ corresponding to different occupations $n_j$ at $M$ sites. If the light probe is in a coherent state and the atoms in a Fock state, the scattered light will also be coherent with an amplitude $\alpha_{\bm{n}}$ \cite{mekhovPRL2009,mekhovPRA2009,mekhovLP2010,mekhovLP2011} dependent on the particular matter configuration: $\alpha_{\bm{n}}=C\bra{\bm{n}} \hat{D}\ket{\bm{n}}$, where $\hat{D}=\sum_j u^*_{\mathrm{out}}(\bm{r}_j)u_{\mathrm{in}}(\bm{r}_j)\hat{n}_j$ sums density-dependent contributions from illuminated sites, $u(\bm{r})$ are the mode functions of probe and scattered light, and $C$ is the Rayleigh scattering coefficient into a cavity or free space \cite{Kozlowski2014}. Due to the linearity of quantum mechanics, since the general atomic state is in a superposition of Fock states, the light and matter become entangled, with joint state $\ket{\psi}= \sum_{\{\bm{n}\}} c^0_{\bm{n}}\ket{\bm{n}}\ket{\alpha_{\bm{n}}}$.

When light is scattered into a cavity with a decay rate $\kappa$, detection of the escaped photons alters the probability amplitudes $c_{\bm{n}}$ \cite{mekhov2012}:  $c_{\bm{n}}(m,t) =\alpha_{\bm{n}}^me^{-|\alpha_{\bm{n}}|^2\kappa t}c^0_{\bm{n}}/\mathcal{N}$ ($\mathcal{N}$ is the normalization), the first factor reflecting $m$ quantum jumps (photons detected) and the second the non-Hermitian evolution during a time $t$. The measurement hence changes the state of both the light and matter, this being the measurement back-action. While light amplitude and phase is measured, the distribution of $\alpha_{\bm{n}}$ is narrowed and the state is gradually projected towards terms with only one $\alpha$ (squeezing below the standard limit is not required). With continuous measurement light is pinned by the quantum Zeno effect, and the atoms will undergo Zeno dynamics \cite{facchi2008, schafer2014, patil2014}, which is constrained such that they may only evolve within the region of Hilbert space with configurations $\ket{\bm{n}}$ corresponding to the measured $\alpha$. Thus, the result of coherent collective scattering strikingly contrasts the outcome of incoherent light scattering, where atoms are localized to a mixed state and coherence is destroyed \cite{PichlerDaley2010}. Such controlled dynamics have application in quantum simulations \cite{stannigel2014}.

If both modes are travelling waves, $\hat{D}=\sum_m e^{im\delta}\hat{n}_m$, where $\delta=(\bm{k}_{\mathrm{out}}-\bm{k}_{\mathrm{in}})\bm{d}$ (for two wave vectors $\bm{k}$ and lattice vector $\bm{d}$). This is a consequence of diffraction \cite{mekhovLP2009}: depending on the angle, the light will show diffraction maxima ($\delta=0, 2\pi,..$) and minima in between. Thus, the choice of $\delta$ (via angles or light frequencies) determines the states to which light can be projected, and hence the corresponding atomic configurations. This choice thus controls the region of Hilbert space to which atomic dynamics is restricted during continuous light measurement. 

Importantly, when $\delta=2\pi/R$ ($R\in \mathbb{Z}$), the atoms at sites $j+mR$ scatter light with the same phase and amplitude, and are therefore indistinguishable to light scattering. This crucially gives rise to $R$ spatial atomic modes: the atoms at indistinguishable sites belong to the same mode, while different atomic modes scatter light with different phases. Physically, $\delta=2\pi/R$ corresponds to the angles of multiple diffraction minima and small number of maxima. For example, for $\delta=2\pi$ (diffraction maximum) one mode ($R=1$) is formed as all atoms scatter light in phase, $\hat{D}=\hat{N}_K$ is the atom number operator for all $K$ illuminated sites (the remaining $M-K$ non-illuminated sites can be considered as an additional mode).  For $\delta=\pi$ (diffraction minimum for orthogonal light waves), two modes are generated ($R=2$) as the atoms at neighboring sites scatter light with opposite phase:  $\hat{D}=\sum_m (-1)^m\hat{n}_m= \hat{N}_\mathrm{even}-\hat{N}_\mathrm{odd}$ gives the number difference between even and odd sites. For other diffraction minima, more spatially overlapping modes are generated as shown in \figsref{figmodel}(b,c) for $R=3$ modes. 

Multiple operators $\hat{D}$ can be measured. It is possible to fully characterize $R$ modes by measuring all of $\delta=2\pi m/R,$ $m\in\left[0,(R-1)/2\right]\cap\mathbb{Z}$, with the $r$th mode composed of sites $j$ satisfying $j$mod$R=r$. The measured operators can be written in the form of an invertible Vandermonde matrix \cite{macon1958}. The inverse then reveals the occupation numbers for each mode (one has a system of linear equations to determine all atom numbers). Ultimately, for $R=M$, this leads to the determination of atom numbers at all lattice sites $n_j$ without the requirement of single-site resolution \cite{GreinerNature2009,BlochNature2011,Ueda2014}. In this case, the quantum measurement will project the state to a single multisite Fock state with well defined atom number at all sites $\ket{n_1,..n_M}$.  This is a direct multimode analogy of preparation of a photon Fock state in a single cavity \cite{HarocheBook}. After this state is achieved, the quantum dynamics under continuous measurement is usually finished.

Although such effective single-site access is useful, here we focus on essentially many-body states. Importantly, even after the atom number in the modes is defined, the modes are still given by superpositions of Fock states, and thus quantum dynamics is not extinct, in contrast to the single-cavity QED case \cite{HarocheBook}. For example, in \figref{figmodel}(c), all three states correspond to the same $\alpha$ and thus are protected in a superposition. The Zeno effect prevents the mode atom numbers from changing, and thus prevents interactions between modes, but not within modes. This results in multiple `virtual' lattices on a single physical lattice. By forgoing measurements with certain $\delta$, the restriction on the interaction between modes is partially lifted. This control over the interaction between modes allows for engineering of desired dynamics.

First, we show how an atomic state akin to photonic PDC state \cite{gerry2005} ($\ket{\psi_\mathrm{PDC}}= \sum_n c_n\ket{n}\ket{n}$) can be realized by measurement, and readily generalized to a multimode case. The initial state is a superfluid (SF) delocalized over all sites. For a large lattice, this can be approximated by the Gutzwiller (mean field) ansatz \cite{lewenstein2012} with a product over all sites, where each site is in a coherent state: $\ket{\psi}=\bigotimes_{i=1}^M \sum_n e^{-\nu/2}\nu^{n/2}/\sqrt{n!}\ket{n}_i$, where $\nu$ is the lattice filling factor. This state can be prepared with an external phase reference \cite{bartlett2007}. Measuring amplitude and phase for $\delta=\pi$ and either post-selecting or using feedback (modifying the trapping potential) \cite{ivanov2014} to get $\langle\hat{D}\rangle_{\delta=\pi}=0$, we project to the state
\begin{equation}
\label{eqpdcpostselect}
\ket{\psi}=\frac{1}{\mathcal{N}}\realsum_n\frac{e^{-\lambda}}{n!}\lambda^n \ket{n}\ket{n},
\end{equation}
where $\lambda=\nu K/R$ is the average initial occupation number of each mode (here $R=2$). The two modes are defined as odd and even sites. Note that while post-selection or feedback is needed to get equal mode occupation, the measurement will deterministically project to a state with a fixed difference in occupation $\Delta N$:
\begin{equation}
\label{eqpdcdeterministic}
\ket{\psi}=\frac{1}{\mathcal{N}}\realsum_n\frac{e^{-\lambda}}{\sqrt{n!}\sqrt{n+\Delta N!}}\lambda^{n+\Delta N/2} \ket{n}\ket{n+\Delta N}.
\end{equation}
For large $N$, $\Delta N$ will become vanishingly small compared to the average number ($\Delta N\leq\mathcal{O}(\langle N\rangle^{1/2})$ and thus $\Delta N\ll\lambda$), hence the states (\ref{eqpdcpostselect}) and (\ref{eqpdcdeterministic}) exhibit many similar properties, such as their entanglement, and post-selection and feedback are not strictly necessary. 

We now generalize the procedure to several modes. By measuring all $\hat{D}$ operators required to characterize $R$ modes, excluding $\delta=0$ (which measures the total atom number), and again post-selecting equal numbers in all modes, we obtain the state
\begin{equation}
\label{eqmpdc}
\ket{\psi}=\frac{1}{\mathcal{N}}\realsum_n\left(\frac{e^{-\lambda}\lambda^n}{n!}\right)^{R/2}\ket{n}^{\otimes R}.
\end{equation}
It is truly multimode, and has genuine multipartite entanglement (that is, any possible bipartitioning of modes shows non-zero entanglement \cite{horodecki2009}). This entanglement can be expressed by the entanglement entropy  \cite{amico2008}; the von Neumann entropy of the reduced density matrix of one of the subsystems (identical for the choice of either subsystem) $E(\ket{\psi}_{AB})=S(\rho_A)=-\mathrm{Tr}(\rho_A\log_2\rho_A).$ For all bipartitionings, if  $\lambda\gg1$, $E=(1/2)\log_2\left(2\pi e {\lambda}/{R}\right).$ As with the two mode case, even without post-selection, the state will still deterministically be projected to one with fixed number differences between each mode, and share similar properties to the `ideal' case (\ref{eqmpdc}).

In quantum optics, simple multimode PDC or four-wave mixing produce multipartite entanglement that is not genuine (entanglement exists between mode pairs, but not between all of them) \cite{Boyer2014}, as photons are produced in pairs, while higher nonlinearities are challenging to achieve.  In contrast, our system produces a kind of genuinely entangled multimode squeezed state which is generally non-Gaussian. In optics, similar continuous variable (CV) Gaussian-like states are obtained using multiple beam splitters, which complicates the scaling to many modes \cite{BraunsteinRMP}, or frequency combs \cite{Averchenko2014}. The atom-optics system we suggest here may provide advantages using the mode entanglement of quantum matter fields.

This method can also be used to create states similar to generalized squeezed states \cite{braunstein1987}, which in optics are expected to be formed from the highly nonlinear process described by the Hamiltonian $H=ga^k+g^*(a^\dagger)^k$. By taking the multimode case above (with $R$ modes taking the place of the $k$-photon process), and then lowering the lattice potential between the modes (but leaving a global trap), the atoms will behave as a single mode, with the state $\ket{\psi}=\sum_n c_n \ket{N_0+nR}$, where $N_0$ and $c_n$ depend on the measured light. This generalizes squeezed vacuums containing even numbers of photons (pairs) to triplets, quadruplets, etc. for increasing $R$.

\begin{figure}
\centering
\includegraphics[width=\linewidth]{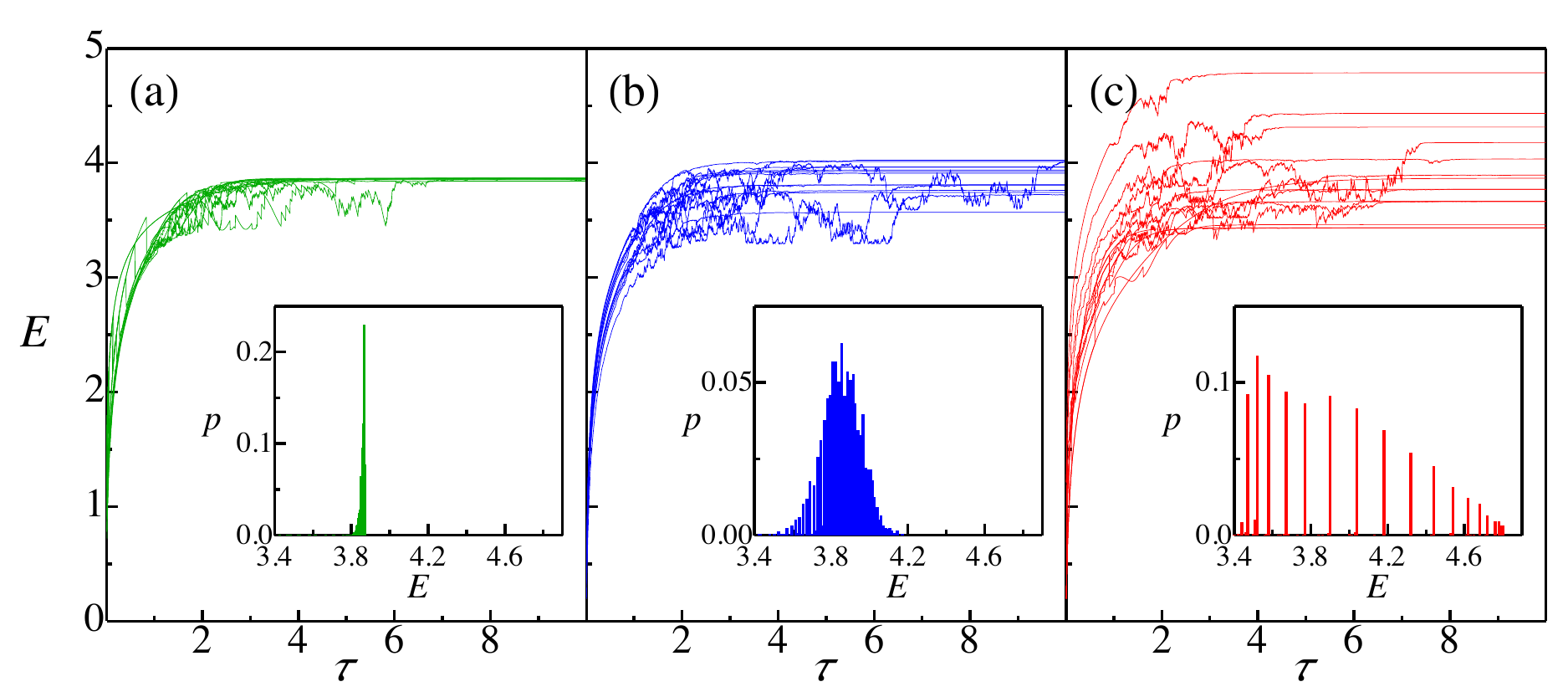}
\caption{Quantum trajectories for the growth of entanglement between two modes where the detected variables are (a) atom number difference in diffraction minimum; (b) total atom number in diffraction maximum; (c) absolute value of atom number difference in minimum. Insets show the final entanglement distribution functions. $\langle N\rangle$=50, $\tau=2|C|^2\kappa t$.}
\label{figtrajectories}
\end{figure}

Measurement at different angles results in other states. A case $\delta=0$ (diffraction maximum) reveals the total number of atoms illuminated, and thus projects to the ``fixed atom-number SF" or multimode generalization of ``spin coherent state". Using several such measurements, or in combination with measurements for other $\delta$ [e.g., 
measuring the total atom number in (\ref{eqpdcpostselect}) - (\ref{eqmpdc})], one prepares a product state of several SFs with $N_i$ atoms: $\ket{\psi}=\bigotimes_{i=1}^R\ket{N_i}_i$ in the mode basis. This corresponds to the multimode generalization of Dicke states if written in the symmetrized particle basis. Note that those SF modes may be non-continuous in space, e.g. one SF can occupy each third site (\figref{figmodel}), which in time-of-flight measurements would be revealed as the period change in the matter-wave interference.

Another interesting case is when for $\delta=\pi$ the amplitude of $\alpha$ is measured, but not the phase. For two modes, this gives the magnitude of the number difference between the two modes $|\Delta N|$, but not its sign. This results in a superposition of cat states
$\sum_n c_n(\ket{n,n+\Delta N}\ket{\alpha_{\Delta N}} \pm \ket{n+\Delta N,n}\ket{-\alpha_{\Delta N}})/\sqrt{2}.$ If the light is turned off, this leads to the atomic cat state, which could be maintained by freezing matter dynamics by other means (e.g. by ramping up the lattice depth). 

In \figref{figtrajectories}, we present quantum trajectories of the evolution of entanglement entropy (using quantum Monte Carlo wave function simulation \cite{CarmichaelBook}) for these three cases: (a) phase-sensitive measurement at the diffraction minimum ($\delta=\pi$); (b) diffraction maximum ($\delta=0$); and (c) phase insensitive measurement at the minimum ($\delta=\pi$) (all for $R=2$ modes). We see that during the measurement, when light-matter entanglement obviously degrades, matter-matter entanglement between initially separable modes is established and grows significantly. While the final average entanglement is similar for all cases, the distribution widths are clearly different (see insets). In \figref{figtrajectories}(a), the distribution is very narrow, as amplitudes of Fock state in Eq. (\ref{eqpdcdeterministic}) only have a weak dependence on $\Delta N$ for the typical $\Delta N \ll \langle N\rangle$. In \figref{figtrajectories}(b), the distribution is broad, as the total atom number measurement projects to SFs with different $N$, and because of the scaling as $\log_2 N^{1/2}$ \cite{simon2002}, the final $E$ depends on the measured atom number. In \figref{figtrajectories}(c), the distribution is even broader as the state is a superposition of cat states. It has the highest probability to be projected to a state with large entanglement even though the average for all scenarios is the same. Importantly, for large $N$, all three widths vanish due to the logarithmic scaling. Thus, the entanglement in effect evolves deterministically, and simulation of a single quantum trajectory will be enough to describe the entanglement evolution, providing a significant numerical simplification.

Next, we show how the multimode structure of matter fields enables entanglement to be established between several spatially separated many-body systems, even if they are initially in Fock states without any phase coherence between them. Importantly, our method does not require any particle subtraction or light-induced particle exchange between the systems, in contrast to previous proposals \cite{Horak1999,RuostekoskiPRA1997}. The idea is to introduce additional sub-modes in the systems, which is possible in our setup. We start with two SFs with fixed atom numbers $N_A$ and $N_B$ (they can be prepared by measuring light at $\delta=0$). We define two new sub-modes within each SF, composed of the odd and even sites, and write the state of each subsystem in the basis $\ket{N_{\mathrm{odd}},N_{\mathrm{even}}}$. Measurement for $\delta=\pi$ across the two subsystems then projects to a state with fixed atom number difference between odd and even sites across the two subsystems: $\ket{\psi}=\sum_k{N_A \choose k}^{1/2}{N_B \choose l(k)}^{1/2}\ket{k,N_A-k}\ket{l(k),N_B-l(k)}/\mathcal{N}$, where $l(k)=(N_A+N_B-\Delta N)/2 -k$, and $\Delta N$ is the measured number difference. The state of subsystem $A$ is then 
$\rho_A=\sum_k {N_A \choose k} {N_B \choose l(k)} \ket{k, N_A-k} \bra{k, N_A-k}/\mathcal{N}^2$. There is thus entanglement between $A$ and $B$ if this reduced state has a non-zero entropy, i.e. ${N_A \choose k} {N_B \choose l(k)}\neq0$ for at least two $k$. This occurs whenever $|\Delta N|\neq (N_A+N_B)$.  Such a scheme for entanglement generation will readily work for multiple initially separable systems.

So far, we have used light detection defined by the on-site ($\hat{n}_i$) density-dependent  operators $\hat{D}$. Note that it is also possible to measure the combinations of conjugate operators $b^\dag_ib_{i+1}$ ($\hat{n}_i=b^\dag_ib_i$) by concentrating the probe light between sites \cite{Kozlowski2014}. Detecting combinations of these operators may enable generation of cluster-like and other states used for QIP \cite{BraunsteinRMP,Sanpera2009}.

\begin{figure}
\includegraphics[width=\linewidth]{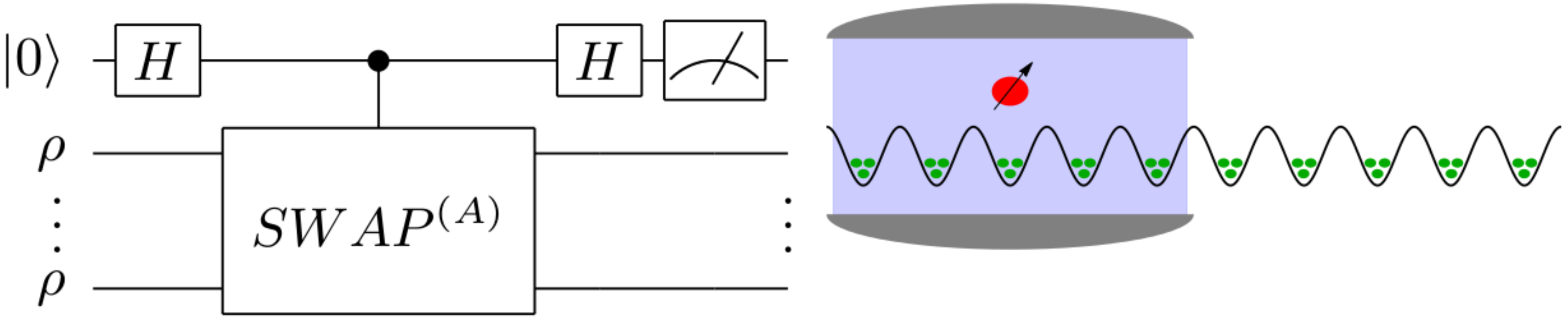}
\caption{Quantum circuit for the measurement  of density matrix moments. Spatially overlapped matter modes are used as multiple copies for the generalized SWAP (permutation) gate. Diagram of the setup for measuring entanglement between illuminated and non-illuminated regions.}
\label{figentanglement}
\end{figure}

We now propose how to exploit the nontrivial spatial overlap (\figref{figmodel}) between the modes for measurement of entanglement in an atomic system. The modes can be used as multiple copies of a system of interest and their overlap enables a straightforward shift from one copy to the neighboring one in space.

Moments $\mathrm{Tr}(\rho^m)$ of the density matrix of a state can be determined by acting generalized SWAP operations on multiple copies of the state conditional on the state of an external qubit, where $m$ is the number of copies \cite{alves2003} (\figref{figentanglement}). Unlike other proposals based on this circuit \cite{daley2012,abanin2012, pichler2013}, ours easily generalizes to arbitrary numbers of dimensions. These moments can be used to find the purity ($m=2$), as well as the R\'{e}nyi entropies $H_m(\rho)=\log_2 (\mathrm{Tr}(\rho^m))/(1-m)$, and, by expanding the logarithm in a Taylor series, the entanglement entropy.

Each virtual lattice is used to support a copy of the system, prepared in an appropriate many-body state of interest. Two internal states $\{\ket{0},\ket{1}\}$ of an impurity atom may be used as the qubit. We allow one of these states to couple to an incident coherent light beam that scatters from the qubit into a cavity, thus the combined state of the qubit and cavity is $(\ket{0}\ket{0}+\ket{1}\ket{\alpha})/\sqrt{2}$. The cavity is positioned such that it also has spatial overlap with one of the sets of subsystems in the lattice, and is then used to drive Rabi transitions in the atoms in this region, so that these atoms are now in another internal state (\figref{figentanglement}). The lattice is chosen such that it is formed from two potentials with equal lattice spacing, one coupling to each of the internal states of the atoms. These potentials are then moved (shifted) by half a site in opposite directions, in the same fashion as a collision gate \cite{jaksch1999}. Essentially, this is possible due to the particular spatial overlap of multiple modes we have proposed here. This transposition of the atom effects the desired SWAP gate, and the probability of detecting the atom in the excited state after a further Rabi $\pi/2$ pulse is proportional to the desired $\mathrm{Tr}(\rho_A^m)$ \cite{alves2003}. This hence determines the entanglement between this subsystem and the (non-illuminated) rest of the system through the von Neumann entropy.

In summary, we have shown how to use quantum measurement of light to construct a multimode structure for ultracold atoms. We have demonstrated how this may be controlled to engineer the states and dynamics of the matter, and provided examples of multimode generalizations of down-conversion, Dicke, SF, and other states. We have also shown how the nontrivial spatial overlap between matter-field modes can be exploited to produce  genuine multipartite entanglement, and used for the measurement of entanglement in quantum gases. Enhanced scattering at a particular angle can be achieved with a cavity. Currently, there are three operating experiments where a Bose-Einstein condensate (BEC) is trapped in a cavity \cite{EsslingerNat2010,HemmerichScience2012,ZimmermannPRL2014}, but without a lattice. There has also been two works where light was scattered from truly ultracold atoms in a lattice, and the measurement object was light \cite{Weitenberg2011,KetterlePRL2011}, without a cavity. The combination of these setups can lead to the realization of our scheme. The primary sources of error we would expect in an experimental implementation would be atom loss due to heating and spontaneous emission, and miscounting of scattered photons due to imperfect detection, both of which increase classical uncertainty leading to reduced purity of the final state. In a recent experiment \cite{patil2014} the Zeno effect was used to localize an atom in a lattice to a single site, demonstrating a simple example of the general ideas discussed here. 

The authors would like to thank Vlatko Vedral, Felix Binder and EPSRC (DTA and EP/I004394/1).

\bibliography{ref}

\section{Supplemental Material}
In the main text we make reference to our setup being described by a generalized Bose-Hubbard Hamiltonian \cite{mekhov2012}, where atoms in an optical lattice scatter light, which is treated as a dynamical quantum variable. Here we provide this Hamiltonian, and give a brief description of the origin of each term.

The full Hamiltonian of the system is given by
$$H=H_L+H_M+H_{LM}.$$

The light mode Hamiltonian
$$H_L=\realsum_l \omega_l a^\dagger_l a_l $$
contains terms for the energy in the light modes. Photons in mode $l$ with frequency $\omega_l$ and mode function $u_l(\bm{r})$ are created (annihilated) by boson operators $a_l^\dagger$ ($a_l$). Note that here we have adopted natural units, such that $\hbar=1$.

The matter behaves according to the Bose-Hubbard Hamiltonian
$$H_M=-t\realsum_{<i,j>}(b_i^\dagger b_j+b_j^\dagger b_i) + U\realsum_ib^\dagger_ib^\dagger_ib_ib_i,$$
where the first term represents tunneling between sites (the sum is over nearest neighbours), and the second $s$-wave scattering between atoms occupying the same site. We consider bosonic matter modes created (annihilated) at site $i$ by $b^\dagger_i$ ($b_i$). We will treat the trapping potential classically, so that we may focus on the quantum nature of the external light fields.

The last term describes the light-matter interaction. It is derived from a many-atom generalisation of the Jaynes-Cummings model with adiabatic elimination of the excited states.
$$H_{LM}=\frac{1}{\Delta_a}\realsum_{l,m}\realsum_{i,j}J^{lm}_{ij}g_lg_ma_l^\dagger a_mb_i^\dagger b_j,$$
where
$$J_{ij}^{lm}=\int\d\bm{r}w(\bm{r}-\bm{r}_i)u^*_l(\bm{r})u_m(\bm{r})w(\bm{r}-\bm{r}_j),$$ 
and $w(\bm{r})$ are the Wannier functions of the lattice. The atom-light detuning $\Delta_a=\omega-\omega_a$, where $\omega_a$ is the frequency of the atomic transition, and $\omega$ is some central frequency set by the adiabatic elimination, and $g_l$ are the light-atom coupling strengths. This term governs the new properties present in the fully-quantum treatment. While in a classical treatment of the light this term can still allow control over the range, magnitude and phase of tunnelling terms through the mode functions $u(\bm{r})$, in the quantum case the light mode operators $a_l$ are also present, and this allows the interaction to feed back into the light modes, elevating light to a dynamical component of the system.

\end{document}